\documentclass[preprint,12pt,authoryear]{elsarticle}

\usepackage{amssymb}
\usepackage{amsmath}
\usepackage{graphicx}
\usepackage{longtable}
\usepackage{multirow}
\usepackage{makecell}
\usepackage{float}
\usepackage{color}
\usepackage{comment}
\usepackage{booktabs}

\journal{}

\begin{document}

\begin{frontmatter}



\title{Advancement of Circular Economy Through Interdisciplinary Collaboration: A Bibliometric Approach} 


\author[inst1]{Keita Nishimoto}
\author[inst1]{Koji Kimita}
\author[inst1]{Shinsuke Murakami}
\author[inst1]{Yin Long}
\author[inst1]{Kimitaka Asatani}
\author[inst1]{Ichiro Sakata}

\affiliation[inst1]{organization={The University of Tokyo},
            addressline={7-3-1 Hongo}, 
            city={Bunkyo-ku},
            postcode={113-8654}, 
            state={Tokyo},
            country={Japan}}

\begin{abstract}
Since the European Union introduced its Circular Economy (CE) Action Plan in 2015, CE research has expanded rapidly. However, the structure of this emerging field—both in terms of its constituent disciplines and researcher dynamics—remains poorly understood. To address this gap, we analyze over 25,000 CE-related publications from Scopus by combining conventional bibliometric approaches with advanced machine learning techniques, including text embeddings and clustering. This hybrid method enables both a macro-level mapping of research domains and a micro-level investigation of individual researchers’ disciplinary backgrounds and collaborations.

We classify CE research into 16 distinct clusters, identifying the original disciplines of researchers and visualizing patterns of interdisciplinary collaboration. Building on this foundation, we ask: Which CE-related research domains receive the most attention in academic and policy contexts? And how are different types of interdisciplinary collaboration associated with research impact?

Our findings show that research in business and management attracts substantial academic and policy attention, while engineering research—though less visible—tends to achieve higher funding success. This suggests a positive dynamic in which the former draws attention to CE issues and the latter secures the economic resources necessary to realize them.

We further demonstrate that CE papers co-authored by researchers from different disciplines tend to show higher research impact than intradisciplinary work. Qualitative case analyses also highlight this tendency. Centered particularly on collaborations between business-oriented and engineering-oriented disciplines, our findings underscore the importance of interdisciplinary efforts in CE research and offer insights for guiding future cross-disciplinary engagement in the field.
\end{abstract}

\begin{graphicalabstract}
\end{graphicalabstract}

\begin{highlights}
\item Analyzed 25,000+ papers to reveal CE research fields and key topics
\item Uncovered field shifts and interdisciplinary ties among CE researchers
\item Interdisciplinary papers show higher impact than single-discipline ones
\end{highlights}


\end{frontmatter}



\section{Introduction}
\label{sec:introduction}
Circular Economy (CE) gained momentum after the European Union announced 54 specific actions in 2015 to promote a circular economy.
Furthermore, in 2020, the EU released a new Circular Economy Action Plan, accelerating research and social implementation efforts toward its realization.

Many researchers have pointed out the necessity of an interdisciplinary approach to realize CE \citep{lieder2016towards, pomponi2017circular, marra2018knowledge}.
For example, in product design, it is necessary to consider not only structural aspects such as design for longevity and design for disassembly, but also business models such as Product-Service Systems \citep{tukker2004eight, kjaer2019product}.
To implement such business models, it is important to transform organizational capabilities not only in product development but also in areas like sales and service provision, thus requiring knowledge in organization and management \citep
{kimita2022servitization}.
In addition, new business models such as sharing and pay-per-use must take into account consumer acceptance and behavioral change, making insights from marketing and the social sciences essential \citep
{edbring2016exploring, coderoni2020sustainable}.
Moreover, to disseminate these products and business models, in addition to bottom-up research at the firm level, top-down research on legislation and policy is needed to build supporting infrastructure and tax systems \citep
{lieder2016towards, pomponi2017circular}.

As existing review articles have pointed out \citep{alcalde-calonge_evolution_2022}, the concept of Circular Economy encompasses a wide range of research fields.
By fostering interdisciplinary collaboration among diverse research domains, it is possible to more effectively implement the CE concept.
This paper analyzes the role of interdisciplinary collaboration in promoting a circular economy using a bibliometric approach.
Specifically, we aim to answer the following research questions:

\begin{itemize}
\item RQ1. Among the research fields constituting the Circular Economy, which ones are receiving attention and citations from academic and policy perspectives?
\item RQ2. Which types of interdisciplinary collaborations are emerging, and how are such collaborations related to research impact?
\end{itemize}

This study not only presents the current status and trends of the entire field from a macro perspective, but also conducts a micro-level analysis of the collaborations taking place.

The structure of this paper is as follows: Chapter 2 reviews existing literature on Circular Economy and bibliometric analyses, highlighting the novelty and significance of this study.
Chapter 3 describes the dataset and methods used in this paper.
In Chapter 4, we first address RQ1 by presenting basic analyses of the research fields that constitute Circular Economy (Section 4.1).
Then, by comparing periods before and after the proposal of the first Circular Economy Action Plan, we identify the interdisciplinary trajectories of researchers and collaboration to answer RQ2 (Section 4.2).
In this section, we select several researchers and trace their career histories and co-authorships to concretely illustrate the types of collaborations that are occurring.
Finally, in Chapter 5, we summarize our findings and suggest directions for future collaboration in the Circular Economy field.

\section{Literature Review}
\label{sec:literature_review}

Although the concept of CE itself has existed since the 1960s \citep{blomsma_emergence_2017}, as will be shown later, the number of publications increased rapidly after the EU introduced its action plan in 2015.
Therefore, this section focuses primarily on research conducted after 2015.

While CE has been rapidly expanding, it remains a concept encompassing a wide range of fields and definitions \citep{kirchherr_conceptualizing_2017, alcalde-calonge_evolution_2022}.
As such, many studies have aimed to clarify the definitions included in this emerging concept, as well as the major research topics and keywords.
These studies can be broadly categorized into qualitative \citep{tukker_product_2015, lieder2016towards} and quantitative analyses \citep{nobre_scientific_2017, turkeli_circular_2018, ruiz-real_worldwide_2018, camon_luis_circular_2020, ranjbari_two_2021, dragomir_state_2024}.
The bibliometric approach adopted in this study falls under the latter, and the following review will focus on quantitative analyses (for a comprehensive review of the former, see \citep{goyal_circular_2021}).

Bibliometric approaches analyze the evolution of scientific knowledge in CE from various perspectives, such as publication patterns, citation networks, and co-authorship networks.
Türkeli et al. \citep{turkeli_circular_2018} conducted bibliometric and network analyses, along with surveys of researchers, on 412 CE papers from Web of Science (WoS) and 1290 from Scopus, focusing mainly on the EU and China.
They tracked the key authors, institutions, cities, cited works, and funding sources in CE-related scientific production in these regions.
They revealed that, as of 2016, China and the EU had the highest number of CE publications and were also each other's major co-authorship partners.
A notable feature of their study is the city-level analysis, which showed that Beijing, Shenyang, Dalian, and Shanghai were the most active cities for CE research in China, while Delft was the most active in the EU.

Nobre \citep{nobre_scientific_2017} focused on IoT and big data technologies as tools for implementing CE in real-world contexts and examined how these technologies are applied in CE-related research.
They extracted 70 CE-related papers on big data and IoT, and conducted content analysis and social network analysis.
Their findings indicated that the United States and China showed the most interest in this field.
On the other hand, they pointed out a gap between the practical use of IoT/big data in the private/industrial sector and the academic sector, which remains at the stage of ``imagining the possibilities.''

Around the same time, Ruiz-Real \citep{ruiz-real_worldwide_2018} conducted a mid-scale bibliometric analysis.
They analyzed 743 CE-related papers extracted from WoS, conducting a basic bibliometric analysis of publication trends and publication numbers by country.

Since 2020, as the cumulative number of CE-related papers has increased, there has been a rise in mid- to large-scale bibliometric studies using thousands of extracted papers.
Camón Luis et al. \citep{camon_luis_circular_2020} conducted bibliometric analyses, including publication trends, citation analysis, and keyword analysis, based on 3391 and 1901 CE papers published between 2016 and 2019 from WoS and Scopus, respectively.
Additionally, using a strategic diagram approach, they found that while early CE research focused on production aspects, recent studies have shifted toward the influence of CE on corporate organizations and organizational structures.
Alnajem et al. (2020) conducted a study aimed at comprehensively capturing the dynamic growth of CE, analyzing 214 papers published between 2009 and 2018 by dividing them into three phases: 2009–2013, 2014–2016, and 2017–2018.
In addition to general bibliometric analysis, they performed multiple correspondence analysis (MCA) to reveal the structure of keywords used in CE.
According to their analysis, CE-related keywords can be broadly categorized into two groups: those related to CE practices such as life cycle assessment, industrial ecology, technologies, and waste management, and those related to CE logistics, such as supply chain and reverse logistics.

Among more recent studies, Ranjbari et al. \citep{ranjbari_two_2021} conducted a comprehensive analysis of waste management (WM) research in the CE context over the past two decades (2001–2020).
Using 962 papers extracted from WoS, they identified prominent research themes in CE-related waste management: (1) bio-based WM; (2) CE transition; (3) electronic waste; (4) municipal solid waste; (5) environmental impacts and lifecycle assessment; (6) plastic waste; and (7) construction and demolition WM.

One of the most recent bibliometric studies is by Dragomir et al. \citep{dragomir_state_2024}.
They analyzed 13553 CE-related papers published between 2006 and 2023 by research institutions in EU member states, addressing two research questions:
RQ1: What is the CE-related scientific output of research institutions within the EU?
RQ2: Which are the CE-related research interests of authors affiliated with institutions in each EU country?
For RQ1, they noted that many of the most-cited CE papers are review articles, and found that Italy had the highest research productivity and citation count within the EU.
For RQ2, they used co-word analysis to identify five research clusters:
C1: Sustainable development and life cycle assessment,
C2: Biomass production and waste valorization,
C3: Materials and recycling,
C4: Wastewater treatment and environmental pollution,
C5: Carbon emissions reduction and energy recovery.
They demonstrated that these topics align with the keywords of the EU's 2020 Circular Economy Action Plan.

Many other studies have also investigated the state of the Circular Economy within specific research domains. For example, Saha et al. \citep{SAHA2025145427} conducted a review of the Circular Economy in healthcare, analyzing 2589 papers retrieved from Scopus, while Neri et al. \citep{neri2025relationship} reviewed the relationship between Digital Technologies and the Circular Economy based on 4379 papers from Scopus.

Summarizing the existing literature, many studies focus on tens to thousands of papers or specialize in individual domains that constitute the CE, and even studies that analyze tens of thousands of papers, such as Dragomir et al. \citep{dragomir_state_2024}, are regionally limited.
The present study distinguishes itself by attempting to link these various domains, whose dynamics have typically been analyzed separately. Specifically, based on a large dataset of over 25,000 papers and employing machine learning techniques, it (1) classifies the research domains constituting the CE into 16 fine-grained categories, (2) maps the landscape of attention to these domains from both academia and, importantly, from policy documents—which have rarely been examined—and (3) quantitatively elucidates the dynamics of inter-domain collaboration and its relationship to research impact.

\section{Datasets}
\label{sec:data_and_methodology}

\begin{figure}[h]
\centering
\includegraphics[width=\linewidth]{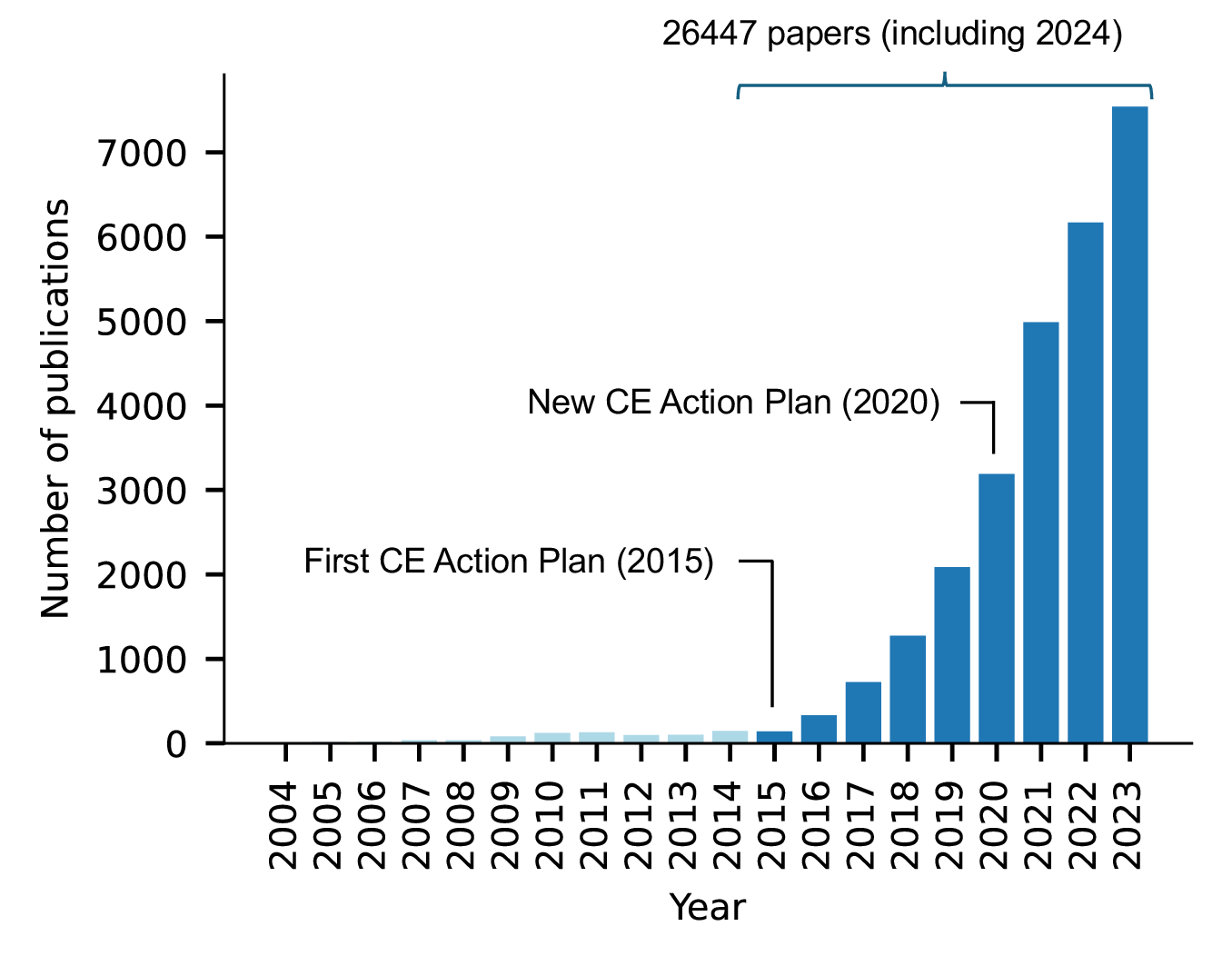}
\caption{Annual publication trends of Circular Economy papers.}
\label{fig1}
\end{figure}

In this study, we use a dataset of academic papers related to the Circular Economy and a dataset of policy documents.
For the academic paper dataset, we extracted 27231 papers from Elsevier's Scopus dataset that include the term ``Circular Economy'' or ``circular economy'' in the title, abstract, or keywords.
Among these, we focus on 26447 papers published between 2015 and 2024, following the announcement of the EU's first action plan (hereinafter referred to as CE papers).
Papers with empty abstracts were excluded from the dataset.
In addition to information such as article titles, author names, journal names, and publication years, Scopus also provides data on citation counts and details regarding research grants acknowledged in each publication.
Figure~\ref{fig1} shows the annual trends in the number of publications. Note that the year 2024 is not included in the figure, as only 9 publications were recorded due to the limited data collection period in Scopus.
Although research on the Circular Economy has been conducted since the 2000s, this study focuses on the period after 2015, when the first action plan was introduced and the number of publications increased.

For the policy document dataset, we extracted 20726 policy documents from Overton's policy citation database that include the topic ``Circular economy'' and contain the phrase ``circular economy'' (retrieved in September 2024).
We refer to these as ``CE-related policy documents,'' and our analysis focuses on those that cite CE papers.
Among the academic papers cited by CE-related policy documents, 33702 papers were successfully matched with DOIs, of which 1972 were CE papers.
These 1972 papers were cited by 1518 policy documents, which we use in the policy-related analysis.

\section{Results}
\label{sec:results}
\subsection{Basic analysis}
\label{subsec:basic_analysis}
\subsubsection{Research fields constituting Circular Economy}
\label{subsec:research_fields}

\begin{figure}[h]
\centering
\includegraphics[width=\linewidth]{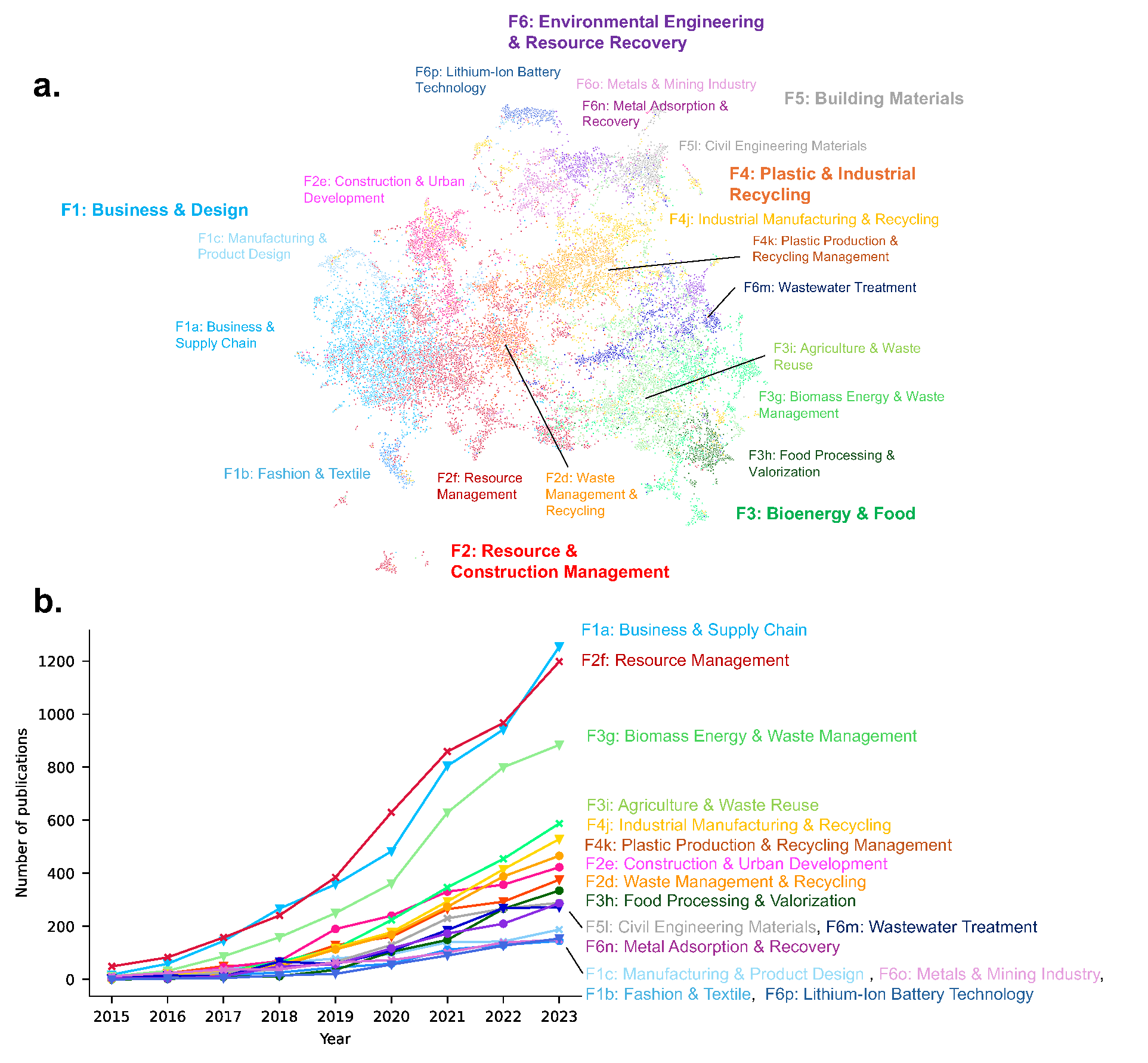}
\caption{(a) Two-dimensional representation of the embedded abstracts of CE papers. (b) Publication trends over time for each research field.}
\label{fig2}
\end{figure}

As shown in previous research \citep{alcalde-calonge_evolution_2022}, the Circular Economy consists of multiple research fields.
Recent advances in language models have made it possible to generate semantic vectors from paper abstracts, capturing their meaning in a form suitable for computational analysis. 
By clustering these vectors, the underlying structure of research fields can be effectively identified \citep{nagao2024researcher}.
Here, we input the abstracts of CE papers into OpenAI's text-embedding-3-small model, and performed hierarchical clustering using the 1536-dimensional semantic vectors it produced, applying Ward's method \citep{ward1963hierarchical} to classify them into six major and sixteen minor research field.
Each research field was labeled by analyzing the top 20 Term Frequency–Inverse Document Frequency (TF-IDF) for each cluster, along with a manual review of representative paper abstracts and titles.
From the top 20 words, terms directly referring to the Circular Economy—such as circular, circularity, circulating, economy—as well as generic terms like abstract, were removed.
The semantic vector space was reduced to two dimensions using UMAP \citep{mcinnes_uniform_2020}, and the results of the 16-category classification were visualized with color coding in Figure~\ref{fig2}(a).
We can observe that neighboring clusters tend to share the same color, indicating that semantically similar abstracts were classified into the same cluster.
Table~\ref{table1} presents the results of the hierarchical classification along with the top 20 TF-IDF terms for each research field.
In addition, Appendix A lists the most-cited paper within each minor category.

Major Research Field F1: Business \& Design includes many papers focusing on business models and product design.
Among its subfields, F1a: Business \& Supply Chain contains the largest number of papers, with numerous studies discussing the concept of the Circular Economy \citep{geissdoerferCircularEconomyNew2017, kirchherrConceptualizingCircularEconomy2017} and proposing business models for transitioning toward it \citep{bockenProductDesignBusiness2016}.
F1b: Fashion \& Textile covers topics such as reuse and recycling of textile products \citep{sandinEnvironmentalImpactTextile2018}, CE initiatives in the textile and apparel industries \citep{jiaCircularEconomyTextile2020}, and sustainable business models in fashion \citep{todeschiniInnovativeSustainableBusiness2017}.
F1c: Manufacturing \& Product Design centers on remanufacturing, including consumer attitudes and decision-making toward remanufactured products \citep{wangConsumerProductKnowledge2016, vanweeldenPavingWayCircular2016, hazenRemanufacturingCircularEconomy2017} and identifying challenges in implementing remanufacturing processes \citep{kurilova-palisaitieneRemanufacturingChallengesPossible2018}.

Major Research Field F2: Resource \& Construction Management encompasses a broad range of topics, including waste, resource management, and urban development.
F2d: Waste Management \& Recycling addresses various waste types, including construction waste \citep{huangConstructionDemolitionWaste2018} and electronic waste (E-waste) \citep{cucchiellaRecyclingWEEEsEconomic2015}.
F2e: Construction \& Urban Development includes frameworks for conducting CE research in the built environment \citep{pomponiCircularEconomyBuilt2017}, and principles for managing demolition waste and construction \citep{galvez-martosConstructionDemolitionWaste2018}.
F2f: Resource Management includes diverse terms related to resources such as waste, energy, food, and resource, with paper topics ranging from green chemistry \citep{sheldonMetricsGreenChemistry2018, sheldonFactor25Years2017} to technological and policy interventions for decarbonization \citep{rissmanTechnologiesPoliciesDecarbonize2020}, and plastic waste management \citep{vanapalliChallengesStrategiesEffective2021}.

Major Research Field F3: Bioenergy \& Food focuses on bioenergy, agriculture, and food waste.
F3g: Biomass Energy \& Waste Management includes discussions on biorefineries using biomass \citep{venkatamohanWasteBiorefineryModels2016}, waste-to-energy \citep{malinauskaiteMunicipalSolidWaste2017}, and energy use from sewage sludge \citep{kacprzakSewageSludgeDisposal2017a}.
F3h: Food Processing \& Valorization contains papers on circular economy approaches utilizing food products, crops, and by-products generated during food processing \citep{romaniHealthEffectsPhenolic2019, faustinoAgroFoodByproductsNew2019, camposManagementFruitIndustrial2020}.
F3i: Agriculture \& Waste Reuse includes many studies on wastewater reuse in irrigation \citep{christouPotentialImplicationsReclaimed2017}, and reuse of waste as fertilizer or feed \citep{chojnackaBiobasedFertilizersPractical2020, grimmMushroomCultivationCircular2018}.

Major Research Field F4: Plastic \& Industrial Recycling centers on polymers and plastics but also includes metals and glass, focusing on their production, disposal, and recycling.
F4j: Industrial Manufacturing \& Recycling covers diverse topics including recycling of carbon fiber polymers \citep{zhangCurrentStatusCarbon2020}, improvements in manufacturing via 3D printing \citep{despeisseUnlockingValueCircular2017}, and algae-based nanomaterial production \citep{khannaAlgaebasedMetallicNanoparticles2019}.
F4k: Plastic Production \& Recycling Management specifically addresses plastic \citep{vollmerMechanicalRecyclingGiving2020}, bioplastics \citep{rosenboomBioplasticsCircularEconomy2022}, and food packaging, including studies on their recycling and health impacts \citep{geuekeFoodPackagingCircular2018, rhodesPlasticPollutionPotential2018}.

Major Research Field F5: Building Materials focuses on the disposal and reuse of construction materials and includes the single minor field (F5l: Building Materials).
While similar to F2e: Construction \& Urban Development, this category is specifically characterized by its focus on materials such as concrete and cement.
Specific studies include cases of reused concrete from demolished buildings \citep{silvaUseRecycledAggregates2019} and the use of waste as binders or cement substitutes \citep{josephUseMunicipalSolid2018}.

Major Research Field F6: Environmental Engineering \& Resource Recovery covers topics such as extraction of metal resources from wastewater and waste, as well as lithium-ion batteries.
F6m: Wastewater Treatment includes research on the water-energy-food nexus \citep{dodoricoGlobalFoodEnergyWaterNexus2018}, water reuse \citep{voulvoulisWaterReuseCircular2018}, and resource recovery from wastewater \citep{puyolResourceRecoveryWastewater2017}.
F6n: Metal Adsorption \& Recovery focuses on extracting and adsorbing metals from wastewater and waste, including studies on chromium \citep{pengRemovalChromiumWastewater2020a}, methylene blue \citep{novaisBiomassFlyAsh2018}, and biosorption \citep{dodsonBioderivedMaterialsGreen2015}.
F6o: Metals \& Mining Industry includes studies on extracting and recycling precious metals from E-waste and mining waste \citep{isildarElectronicWasteSecondary2018, zengUrbanMiningEWaste2018}.
F6p: Lithium-Ion Battery Technology is a specialized subfield covering battery recycling \citep{swainRecoveryRecyclingLithium2017}, especially lithium-ion battery processing \citep{mossaliLithiumionBatteriesCircular2020, makuzaPyrometallurgicalOptionsRecycling2021}.

Figure~\ref{fig2}(b) shows the publication trends over time for each research field.  
There is a variation in publication volume across the fields, with the highest numbers observed in F1a: Business \& Supply Chain and F2f: Resource Management.  
The former includes terms related to business model transitions, such as ``business,'' ``framework,'' and ``transition,'' as well as words commonly found in review articles, such as ``literature'' and ``concept.''  
Indeed, the most cited paper in this field is ``The Circular Economy – A new sustainability paradigm?'' \citep{geissdoerferCircularEconomyNew2017} (see Appendix A), which aims to clarify the concept of the Circular Economy through bibliometric methods.
On the other hand, the latter contains a wide range of resource-related terms such as ``waste,'' ``energy,'' ``food,'' and ``resource.''  
For example, the most cited paper in this field is ``The E factor 25 years on: The rise of green chemistry and sustainability'' \citep{sheldonFactor25Years2017} (see Appendix A), which analyzes the impact of green chemistry and the E-factor on resource efficiency and waste minimization.  
Meanwhile, technically specialized fields such as F6o: Metals \& Mining Industry and F6p: Lithium-Ion Battery Technology, under F6: Environmental Engineering \& Resource Recovery, show relatively limited growth in publication volume.  
One possible factor contributing to this discrepancy is the number of researchers entering each field after 2015 (analyzed in detail later).  
It has been revealed that while many researchers have moved into the F1: Business \& Design since 2015, very few have moved into the F6: Environmental Engineering \& Resource Management.

\begin{table}[h]
\centering
\caption{Major and minor classifications of research fields with number of publications and frequently appearing terms within each field.}
\resizebox{\textwidth}{!}{ 
\begin{tabular}{cccccp{10cm}}
\toprule
\textbf{ID(6)} & \textbf{Name(6)} & \textbf{ID(16)} & \textbf{Name(16)} & \textbf{\#(16)} & \textbf{Words} \\
\midrule
\multirow{3}{*}{F1} & \multirow{3}{*}{\makecell{Business \& Design}} & F1a & \makecell{Business \\ \& Supply Chain} & 4334 & business, supply, model, sustainability, product, design, literature, chain, development, value, framework, transition, industry, innovation, implementation, management, approach, performance, manufacturing, concept \\
 &  & F1b & \makecell{Fashion \\ \& Textile} & 538 & textile, fashion, industry, clothing, waste, business, apparel, sustainability, design, production, consumer, chain, consumption, model, cotton, value, product, supply, recycling, sector \\
 &  & F1c & \makecell{Manufacturing \\ \& Product Design} & 782 & remanufacturing, product, repair, design, life, manufacturing, reuse, model, equipment, disassembly, value, system, business, use, case, process, consumer, industry, approach, production \\
\midrule
\multirow{3}{*}{F2} & \multirow{3}{*}{\makecell{Resource \\ \& Construction Management}} & F2d & \makecell{Waste Management \\ \& Recycling} & 1370 & waste, management, collection, recycling, system, msw, model, generation, policy, development, recovery, weee, plastic, disposal, treatment, sector, china, resource, household, environment \\
 &  & F2e & \makecell{Construction \& Urban Development} & 1673 & construction, building, design, waste, city, reuse, material, demolition, industry, development, life, sector, environment, management, use, project, framework, sustainability, approach, cycle \\
 &  & F2f & \makecell{Resource Management} & 4572 & waste, development, energy, food, sustainability, management, model, production, system, business, policy, industry, resource, transition, sector, approach, consumption, use, efficiency, environment \\
\midrule
\multirow{3}{*}{F3} & \multirow{3}{*}{\makecell{Bioenergy \& Food}} & F3g & \makecell{Biomass Energy \\ \& Waste Management} & 3214 & waste, energy, production, biomass, biogas, process, food, carbon, sludge, co, treatment, biochar, digestion, oil, bioeconomy, use, conversion, fuel, gas, wastewater \\
 &  & F3h & \makecell{Food Processing \\ \& Valorization} & 913 & food, extraction, waste, fruit, production, pomace, oil, content, activity, industry, peel, processing, extract, use, valorization, health, grape, source, acid, review \\
 &  & F3i & \makecell{Agriculture \\ \& Waste Reuse} & 1837 & production, soil, food, biomass, waste, wastewater, microalgae, treatment, growth, compost, use, feed, water, protein, plant, content, fertilizer, manure, cultivation, process \\
\midrule
\multirow{2}{*}{F4} & \multirow{2}{*}{\makecell{Plastic \\ \& Industrial Recycling}} & F4j & \makecell{Industrial Manufacturing \\ \& Recycling} & 1646 & waste, material, energy, production, process, use, pv, manufacturing, recycling, life, glass, printing, carbon, strength, industry, polymer, performance, construction, wood, rubber \\
 &  & F4k & \makecell{Plastic Production \\ \& Recycling Management} & 1494 & plastic, packaging, waste, recycling, pet, production, pyrolysis, food, use, material, life, polymer, pollution, chemical, process, polyethylene, environment, management, cycle, quality \\
\midrule
\multirow{1}{*}{F5} & \multirow{1}{*}{\makecell{Building Materials}} & F5l & \makecell{Building Materials} & 1060 & concrete, cement, strength, waste, ash, construction, slag, use, material, aggregate, asphalt, sand, production, binder, replacement, steel, water, industry, glass, performance \\
\midrule
\multirow{4}{*}{F6} & \multirow{4}{*}{\makecell{Environmental Engineering \\ \& Resource Recovery}} & F6m & \makecell{Wastewater Treatment} & 990 & water, wastewater, treatment, recovery, membrane, phosphorus, energy, reuse, management, removal, waste, process, sludge, resource, system, phosphate, use, technology, nitrogen, production \\
 &  & F6n & \makecell{Metal Adsorption \\ \& Recovery} & 914 & adsorption, waste, recovery, process, removal, copper, metal, water, treatment, cu, mg, iron, wastewater, acid, mining, concentration, solution, adsorbent, zn, ph \\
 &  & F6o & \makecell{Metals \\ \& Mining Industry} & 639 & mining, steel, industry, waste, metal, energy, supply, coal, scrap, material, production, recycling, demand, development, recovery, copper, use, process, resource, aluminum \\
 &  & F6p & \makecell{Lithium-Ion \\ Battery Technology} & 471 & battery, libs, lithium, ion, energy, storage, li, recycling, life, vehicle, lib, cathode, ev, cobalt, carbon, process, demand, waste, material, evs \\
\bottomrule
\end{tabular} 
\label{table1}
}
\end{table}

\subsubsection{Scientific/Political attentions to each research field}
This section addresses RQ1: Among the various research fields that constitute Circular Economy, which ones receive the most attention and citations from academic and policy perspectives?

\begin{figure}[h]
\centering
\includegraphics[width=\linewidth]{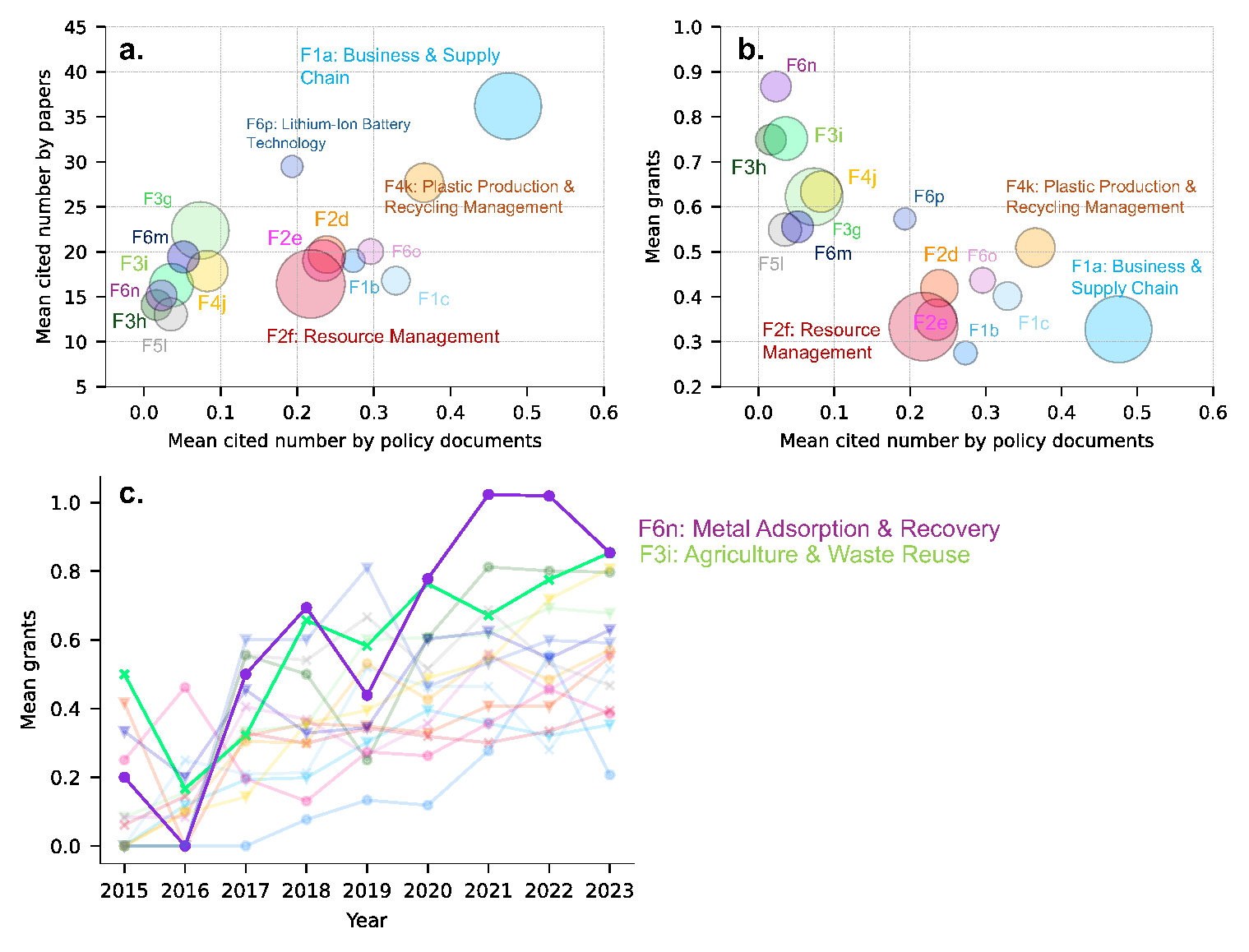}
  \caption{(a) Average number of citations received by papers in each research field from CE-related policy documents (x-axis) and from academic papers (y-axis); bubble size indicates the number of papers in the cluster. (b) Average number of citations from policy documents (x-axis) and average number of grants per paper (y-axis); bubble size indicates the number of papers in the research field. (c) Time series of average grant counts per paper in each research field.}
\label{fig3}
\end{figure}

Figure~\ref{fig3}(a), (b) illustrates the relationships among three indicators: the average number of citations received by CE papers in each research field from other academic papers, the average number of citations from CE-related policy documents, and the average number of grants received per paper. The bubble size indicates the number of papers in the field.

Figure~\ref{fig3}(a) shows the relationship between citations from policy documents and citations from academic papers.  
While there is some variability, a positive correlation is observed between citations from policy and academic sources.  
This suggests that highly impactful papers may influence both academic and policy spheres, and—as pointed out in prior research \citep{yinCoevolutionPolicyScience2021}—academic and policy domains may mutually shape one another's attention, with focus in one domain drawing attention in the other.

Notably, the field that receives the greatest attention from both academic and policy perspectives is F1a: Business \& Supply Chain.  
This subfield contains papers that explain the concept of Circular Economy and propose business models for transitioning toward it.  
It is frequently cited for its contributions to defining and operationalizing the Circular Economy.

Figure~\ref{fig3}(b) shows the relationship between the number of policy citations and the number of grants received by papers in each research field.  
Interestingly, this relationship reveals a negative correlation.  
That is, there appears to be a trade-off between the level of scholarly or policy attention a research field receives and the amount of grant funding it secures.

Research subfields with moderate to high levels of citation but low grant acquisition are predominantly found in Major Fields F1, such as F1a: Business \& Supply Chain and F1c: Manufacturing \& Product Design.  
Conversely, fields in the low citation–high grant category are mainly found in Major Fields F3-F6 (Bioenergy \& Food, Plastic \& Industrial Recycling, and Environmental Engineering \& Resource Management).

The former group frequently proposes conceptual frameworks for a sustainable society, such as business models, fashion systems, and product design, which are likely to capture public attention.  
The latter group addresses concrete technological issues such as biomass energy, food systems, wastewater treatment, offering tangible solutions to existing societal challenges and thereby attracting financial support.
In fact, Figure~\ref{fig3}(c) shows the temporal trends in average grant counts per research field, indicating a sharp increase in grant support after 2016 in the latter fields such as F6n: Metal Adsorption \& Recovery and F3i: Agriculture \& Waste Reuse.  
However, the narrower focus of the latter fields may limit their visibility and broader scholarly or policy attention.

\begin{figure}[h]
\centering
\includegraphics[width=\linewidth]{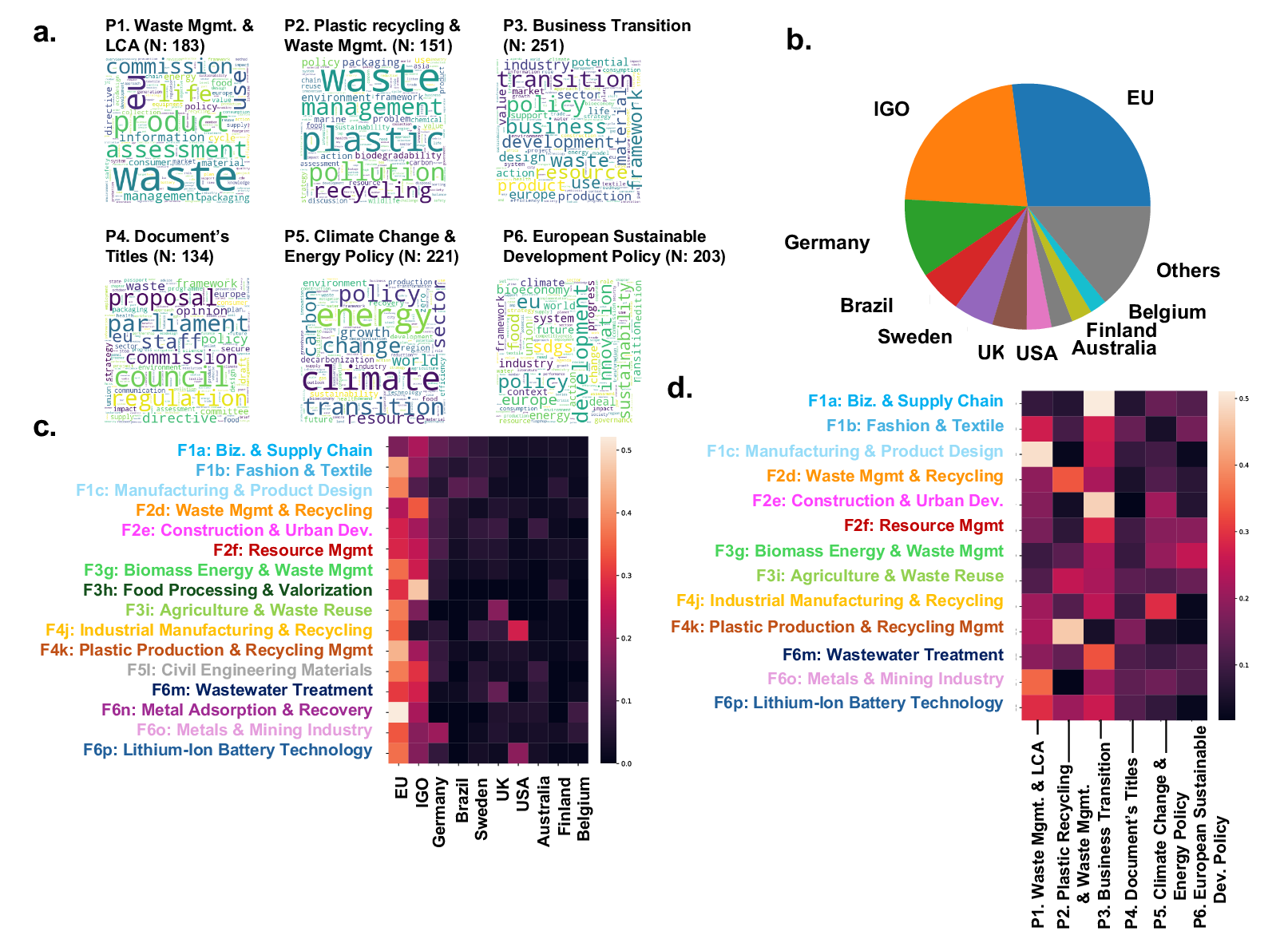}
\caption{(a) Classification results of CE-related policy documents that cite CE papers, (b) Share of countries issuing CE-related policy documents that cite CE papers, (c) Heatmap showing the proportion of citing policy documents by issuers for each research field.  
Specifically, this heatmap shows the total number of citations to CE papers in each research field from policy documents issued by each issuer, normalized within each research field.  Brighter colors indicate higher citation rates, darker colors indicate lower rates. (d) Heatmap showing the proportion of policy topics citing CE papers in each research field.  
Specifically, this heatmap shows the total number of citations to CE papers in each research field from policy documents associated with each policy topic, normalized within each research field.  
Brighter colors indicate higher citation rates, darker colors indicate lower rates. Note that research fields with fewer than 30 total citations from policy documents (F3h, F5l, F6n) are excluded from the index.
}
\label{fig4}
\end{figure}

We next turn our attention to policy document citations—an area not analyzed in previous review studies.  
Figure~\ref{fig4} presents (a) the topic clusters of policy documents citing CE papers, (b) the share of issuing organizations, (c) the share of citations from each issuer by CE research field, and (d) the share of citations from each policy topics by CE research field.  
For panels (a) and (d), to eliminate language bias, we analyzed 1,143 English-language policy documents out of a total of 1518, and classified them into six clusters.  
The classification was based on semantic vectors (1536 dimensions) generated from the abstracts of policy documents in Overton using OpenAI's \texttt{text-embedding-3-small} model, followed by k-means clustering.  
Cluster labels were manually assigned based on the top TF-IDF terms in each cluster.  
Notably, the fourth cluster (P4) contains frequent terms such as \textit{council} and \textit{parliament}.  
This is likely because Overton sometimes records only the titles as abstracts, and many such documents were grouped into this cluster—resulting in title-specific terms being highly weighted.

As shown in Figure~\ref{fig5}(b), about half of all CE paper citations originate from the EU and intergovernmental organizations (IGOs).  
Within the IGO category, most citations come from the OECD and United Nations; within the EU, the Publications Office of the European Union and the Joint Research Centre account for the majority.  
However, panel (c) reveals some variation across research fields.  
For instance, F1a: Business \& Supply Chain receives more citations from IGOs than the EU, F3i: Agriculture \& Waste Reuse is frequently cited by UK-based entities, while F4j: Industrial Manufacturing \& Recycling and F6p: Lithium-Ion Battery Technology are prominently cited by institutions in the USA.

Panel (d) confirms the expected trend that research fields are primarily cited by policy topics with close thematic relevance.  
For example, F1c: Manufacturing \& Product Design is closely linked to Life Cycle Assessment and is often cited by policy documents in P1: Waste Management \& LCA.  
Similarly, strong associations are observed between F4k: Plastic Production \& Recycling Management and P2: Plastic Recycling \& Waste Management, and between F1a: Business \& Supply Chain and both F2e: Construction \& Urban Development and P3: Business Transition.  
Among the six policy topics, P3: Business Transition holds the largest share, suggesting that policy interest in the CE domain is particularly concentrated on business-oriented issues.

\clearpage

\subsection{Interdisciplinary collaboration among researchers}

The preceding analysis has demonstrated that the CE field is composed of numerous distinct research domains.
This raises two key questions: what disciplines did these researchers originally belong to before entering the field of CE research? Furthermore, under the emerging domain of CE, have collaborations emerged between researchers from different disciplines?

To address these questions, we identified active CE researchers—defined as authors who have published three or more CE papers—and extracted a sample of 7658 such individuals.
We then analyzed the papers they published between 2005 and 2014 (i.e., before the announcement of the EU's first CE Action Plan) by generating 1536-dimensional vector representations using the text-embedding-3-small model.
Following the same procedure described in Section~\ref{subsec:research_fields}, we applied hierarchical clustering (Ward's method \citep{ward1963hierarchical}) to these embeddings and assigned each paper to one of six major or sixteen minor categories.
Each researcher was subsequently classified into the cluster corresponding to the most frequent category among their earlier publications.
In cases where a researcher had an equal number of papers in two or more clusters, the final cluster assignment was made at random.
Table~\ref{table2} presents the hierarchically classified disciplines along with the top 20 TF-IDF terms for each cluster.
Discipline labels were assigned by analyzing the top 20 TF-IDF terms for each cluster, along with a manual review of representative paper abstracts and titles.
Note that the disciplines described here reflect the original disciplinary backgrounds of active CE researchers, which differ from the CE research fields (F1--F6) identified earlier in Section~\ref{subsec:research_fields}.

Major Discipline D1: Business \& Supply Chain Management consists of two subfields: D1a: Business Management \& Innovation Studies and D1b: Operation \& Systems Management.
D1a: Business Management \& Innovation Studies includes a wide range of topics such as climate change \citep{moss_next_2010, oneill_new_2014}, supply chains \citep{seuring_literature_2008, sarkis_organizational_2011}, and innovation systems analysis \citep{hekkert_functions_2007}.
D1b: Operation \& Systems Management covers operations research, particularly decision support in supply chains \citep{ho_multi-criteria_2010, behzadian_state_art_2012, chen_fuzzy_2006}, and includes studies related to Product-Service Systems (PSS) \citep{baines_state-art_2007}.

Major Discipline D2: Sustainable Resource Management comprises three subfields: D2c: Environmental \& Agricultural Management, D2d: Sustainable Energy \& Building Systems, and D2e: Sustainable Industrial Development.
D2c: Environmental \& Agricultural Management includes studies on water quality and agriculture, such as sewage treatment using anammox \citep{kartal_sewage_2010}, climate change impacts on crops \citep{rosenzweig_assessing_2014, asseng_uncertainty_2013}, and nitrogen loss in grain yield \citep{ju_reducing_2009}.
D2d: Sustainable Energy \& Building Systems covers solar and wind power \citep{foley_current_2012}, as well as building life-cycle assessments \citep{ortiz_sustainability_2009, cabeza_life_2014}.
Highly cited studies include reviews on thermal storage in solar power \citep{gil_state_2010}, phase change materials in buildings \citep{cabeza_materials_2011}, and Life Cycle Assessment (LCA) of conventional versus electric vehicles \citep{hawkins_comparative_2013}.
D2e: Sustainable Industrial Development spans diverse topics such as climate change, LCA, and waste management.
Key references include Representative Concentration Pathways (RCPs) for greenhouse gas emissions \citep{van_vuuren_representative_2011, meinshausen_rcp_2011}, LCA methodology \citep{finnveden_recent_2009}, health impacts of climate change \citep{costello_managing_2009}, and E-waste research \citep{widmer_global_2005}.

Major Discipline D3: Materials \& Chemical Engineering includes five subfields: D3f: Material Science \& Nanotechnology, D3g: Chemical Reaction \& Catalysis, D3h: Polymer-based Material Science, D3i: Civil Engineering Materials, and D3j: Materials \& Structural Engineering.
D3f: Material Science \& Nanotechnology covers graphene synthesis \citep{hernandez_high-yield_2008}, reviews on graphene properties \citep{huang_graphene-based_2011}, lanthanide hybrid materials \citep{binnemans_lanthanide-based_2009}, and transparent electronics \citep{fortunato_oxide_2012}.
D3g: Chemical Reaction \& Catalysis features research on ionic liquids \citep{smith_deep_2014, hallett_room-temperature_2011} and gold catalysts \citep{hashmi_gold_2006}.
D3h: Polymer-based Material Science includes cellulose nanoparticles \citep{habibi_cellulose_2010}, PolyVinylidene DiFluoride (PVDF) \citep{martins_electroactive_2014}, nanofibers \citep{teo_review_2006}, and biopolymers \citep{vieira_natural-based_2011}.
D3i: Civil Engineering Materials focuses on geopolymer concrete \citep{mclellan_costs_2011, habert_environmental_2011}, microbial-induced carbonate precipitation \citep{de_muynck_microbial_2010}, and hydration of limestone-blended cement \citep{de_weerdt_hydration_2011}.
D3j: Materials \& Structural Engineering includes 3D printing \citep{tekinalp_highly_2014, tymrak_mechanical_2014}, incremental sheet metal forming process with Computer Numerical Control (CNC) \citep{jeswiet_asymmetric_2005}, surface integrity in cutting operations \citep{jawahir_surface_2011}, design of ultra-high performance concrete \citep{yu_mix_2014}, heat generation during metal cutting \citep{abukhshim_heat_2006}, and blast effects on structures \citep{ngo_blast_2007}.

Major Discipline D4: Energy \& Chemical Engineering includes two subfields: D4k: Chemical \& Energy Engineering and D4l: Renewable Energy \& Fuel Production.
D4k: Chemical \& Energy Engineering includes research on fuel cells and green chemistry.
Highly cited works include reviews on microbial fuel cells \citep{logan_microbial_2006, pant_review_2010}, green chemistry \citep{anastas_green_2009}, and the application of ultrasound in food technologies \citep{chematApplicationsUltrasoundFood2011}.
D4l: Renewable Energy \& Fuel Production primarily focuses on biomass-related studies, and also includes CO$_2$ capture and conversion.
Top-cited papers include reviews on biomass fuel synthesis \citep{huber_synthesis_2006}, and pretreatment technologies for ethanol production from lignocellulosic biomass \citep{alvira_pretreatment_2010, taherzadeh_pretreatment_2008}.

Major Discipline D5: Biochemical \& Food Processing consists of two subfields: D5m: Industrial Biotechnology \& Biofuel Production and D5n: Plant Biochemistry \& Food Science.
D5m: Industrial Biotechnology \& Biofuel Production encompasses a wide range of topics related to industrial applications of chemical and biological reactions.
Top-cited studies include enzyme immobilization \citep{sheldon_enzyme_2007, sheldon_enzyme_2013}, reviews on insects for food/feed \citep{van_huis_potential_2013}, industrial uses of laccases \citep{rodriguez_couto_industrial_2006}, and microbial biosurfactant applications and production \citep{banat_microbial_2010}.
D5n: Plant Biochemistry \& Food Science focuses on biochemistry, biomass, and food science, with significant attention to biodiesel production from algae-bacterial systems.
Key references include reviews on the role of reactive oxygen/nitrogen species in cancer \citep{valko_free_2006}, microalgae for biodiesel \citep{mata_microalgae_2010}, and antioxidant/prooxidant systems \citep{carocho_review_2013}.

Major Discipline D6: Wastewater Treatment \& Soil Remediation includes two subfields: D6o: Wastewater \& Waste Management and D6p: Environmental Chemistry \& Remediation.
D6o: Wastewater \& Waste Management focuses on water and air pollution control and wastewater treatment.
Top-cited papers include reviews on biochar for pollutant control \citep{ahmad_biochar_2014}, changes in soil acidity in China \citep{guo_significant_2010}, anaerobic digestion for sludge disposal \citep{appels_principles_2008}, and nitrogen deposition impacts \citep{liu_enhanced_2013}.
D6p: Environmental Chemistry \& Remediation includes studies on chemical impacts on the environment, water purification, and soil remediation.
Top-cited works feature reviews on photocatalytic water treatment \citep{chong_recent_2010}, proton exchange membrane (PEM) water electrolysis \citep{carmo_comprehensive_2013}, effects of antibiotics in aquatic systems \citep{kummerer_antibiotics_2009}, and occurrence/removal of micropollutants \citep{luo_review_2014}.

\subsubsection{Interdisciplinary trajectories of researchers}

\begin{figure}[h]
\centering
\includegraphics[width=\linewidth]{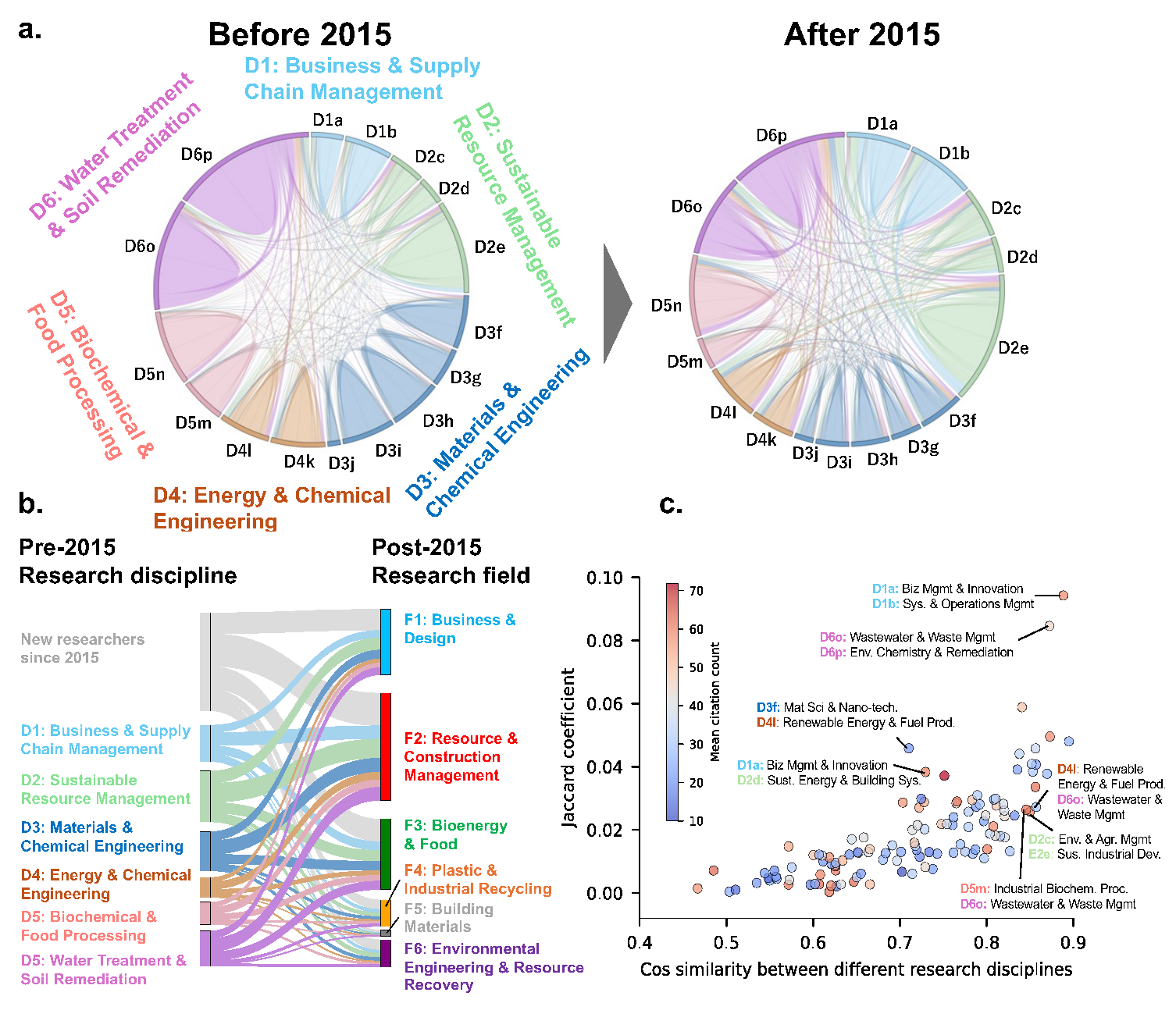}
\caption{(a-Left) Co-authorship volume between disciplines in papers published by active CE researchers during 2005–2014. Each color represents a CE researcher's disciplines, and the thickness of the lines indicates the number of co-authorships.
(a-Right) Co-authorship volume between disciplines in papers published by active CE researchers during 2015–2024. (b) Sankey diagram showing the transition of active CE researchers from their original disciplines during 2005–2014 to their current CE research fields during 2015–2024. The thickness of each colored band represents the number of researchers transitioning from the original disciplines to the research fields. The gray bands represent researchers who newly moved into the CE field after 2015. (c) Jaccard coefficient (y-axis), representing the proportion of co-authorship between disciplines, plotted against the semantic similarity (x-axis) between those disciplines. The color of each plotted circle indicates the average number of citations received by the co-authored papers.}
\label{fig5}
\end{figure}

Figure~\ref{fig5}(b) presents a Sankey diagram showing the migration of active CE researchers from their original major discipline during 2005--2014 to their CE research fields during 2015--2024.
The thickness of each colored band represents the number of researchers who transitioned from the original research discipline to the current research field.
The ``new'' category, shown in gray, represents researchers who published three or more CE papers during 2015--2024 but had no publications during 2005--2014—indicating new entrants to the CE field after the first EU Action Plan announcement.

The diagram suggests that many researchers who were active in other disciplines prior to 2015 moved into the CE field and began contributing to CE-related research.
In particular, Research Field F2: Resource \& Construction Management received a significant influx of researchers.

For example, one such researcher, Dr. Lenny Koh (University of Sheffield), originally conducted research on enterprise resource planning and knowledge management \citep{kohManagingUncertaintyERPcontrolled2006, kohApplicationKnowledgeManagement2005}, which placed her in D1: Business \& Supply Chain Management during 2005--2014.
Over time, she shifted her focus to eco-supply and green supply chains  \citep{koh2007current, koh2009low}, eventually contributing to Circular Economy topics, leading to her current classification under F2: Resource \& Construction Management.

Similarly, Dr. Adolf Acquaye (Khalifa University) initially worked on greenhouse gas emissions in the construction sector \citep{acquayeInputOutputAnalysis2010}, aligning with D2: Sustainable Resource Management during 2005--2014.
In 2012, he co-authored a paper with Dr. Koh on green supply chains \citep{acquaye2012green} and later expanded into broader CE research.
One notable example of their collaboration is the highly cited paper, ``Sustainable supply chain management and the transition towards a circular economy: Evidence and some applications'' \citep{genoveseSustainableSupplyChain2017}, published in 2017 with Dr. Koh and two other co-authors.
This case illustrates a successful interdisciplinary collaboration between researchers from different original disciplines within the CE domain.

Another example involves collaboration among researchers who migrated into F3: Bioenergy \& Food.
Dr. Piergiuseppe Morone (Sapienza University of Rome) originally conducted innovation research on knowledge diffusion networks \citep{morone2004knowledge} and later transitioned through environmental justice \citep{germani2011enforcement} and innovation studies in biorefinery systems \citep{lopolito2011innovation} toward CE-related topics such as the bioeconomy \citep{d2022exploring}.
Dr. Apostolis Koutinas (Agricultural University of Athens) focused initially on biorefinery systems \citep{koutinas2005developing} and later expanded into biodiesel \citep{koutinas2014design} and biopolymer research \citep{koutinas2014valorization}.
Dr. Ioannis Kookos (University of Patras) originally studied control engineering in process systems \citep{kookos2002algorithmic} and later collaborated with Dr. Koutinas on optimization of bioethanol production \citep{arifeen2007process}, continuing into broader research on biofuel production.

Together with four other authors, these three researchers co-authored a CE paper in 2019 titled ``Hybridised sustainability metrics for use in life cycle assessment of bio-based products: resource efficiency and circularity'' \citep{lokesh2020hybridised}.

The Sankey diagram also reveals that research fields with high publication growth—such as F1: Business \& Design and F2: Resource \& Construction Management—have experienced significant researcher inflow, including new entrants.
Conversely, fields with slower growth in publication volume, such as F6: Environmental Engineering \& Resource Recovery, have seen less researcher inflow.
This suggests that differences in researcher migration may explain disparities in publication growth across CE research fields.

\begin{table}[H]
\centering
\caption{Major and minor classifications of active CE researchers' disciplines during 2005–2014, along with frequently appearing terms within each classification.}
\resizebox{\textwidth}{!}{%
\begin{tabular}{cccccp{10cm}}
\toprule
ID(6) & Name(6) & ID(16) & Name(16) & \#(16) & Words \\
\midrule
\multirow{2}{*}{D1} & \multirow{2}{*}{Business \& Supply Chain Management} & D1a & \makecell{Business Management \\ \& Innovation Studies} & 6298 & management, innovation, supply, business, knowledge, chain, development, performance, approach, information, design, literature, model, framework, case, sustainability, technology, process, education, industry \\
 & & D1b & \makecell{Operations \& \\ Systems Management} & 11115 & model, design, system, approach, problem, product, process, decision, management, information, method, algorithm, service, performance, project, time, development, supply, production, case \\
\midrule
\multirow{3}{*}{D2} & \multirow{3}{*}{Sustainable Resource Management} & D2c & \makecell{Environmental \& \\ Agricultural Management} & 7174 & water, soil, model, quality, management, system, wastewater, treatment, use, food, area, production, plant, air, crop, river, wood, irrigation, yield, field \\
 & & D2d & \makecell{Sustainable Energy \\ \& Building Systems} & 4851 & energy, system, building, heat, power, electricity, efficiency, wind, performance, cycle, consumption, model, design, use, air, storage, life, construction, temperature, pv \\
 & & D2e & \makecell{Sustainable \\ Industrial Development} & 8728 & waste, development, management, energy, sustainability, life, product, assessment, cycle, production, policy, system, model, use, impact, process, consumption, approach, design, industry \\
\midrule
\multirow{5}{*}{D3} & \multirow{5}{*}{Materials \& Chemical Engineering} & D3f & \makecell{Material Science \\ \& Nanotechnology} & 8853 & surface, carbon, temperature, electron, tio, structure, microscopy, phase, size, spectroscopy, diffraction, synthesis, material, formation, process, film, deposition, society, method, metal \\
 & & D3g & \makecell{Chemical Reaction \\ \& Catalysis} & 6099 & catalyst, reaction, co, oxidation, activity, synthesis, society, acid, temperature, selectivity, chemistry, surface, hydrogen, pd, conversion, chemical, formation, glycerol, polymerization, pt \\
 & & D3h & \makecell{Polymer-Based \\ Material Science} & 7653 & polymer, matrix, surface, poly, fiber, strength, tissue, release, water, effect, drug, temperature, microscopy, structure, material, pla, cell, content, modulus, morphology \\
 & & D3i & \makecell{Civil Engineering Materials} & 3090 & concrete, cement, strength, waste, ash, water, aggregate, material, mortar, hydration, durability, shrinkage, test, construction, clay, resistance, glass, absorption, slag, portland \\
 & & D3j & \makecell{Materials \& \\ Structural Engineering} & 3726 & steel, process, model, material, design, stress, strength, element, strain, method, surface, crack, concrete, failure, welding, damage, fatigue, temperature, fracture, behaviour \\
\midrule
\multirow{2}{*}{D4} & \multirow{2}{*}{Energy \& Chemical Engineering} & D4k & \makecell{Chemical \& \\ Energy Engineering} & 7445 & membrane, model, water, process, fuel, energy, heat, flow, system, temperature, design, cell, performance, pressure, gas, power, transfer, chemical, work, surface \\
 & & D4l & \makecell{Renewable Energy \\ \& Fuel Production} & 6921 & biomass, energy, production, oil, gas, co, fuel, biodiesel, combustion, process, hydrogen, pyrolysis, coal, carbon, reaction, temperature, gasification, model, conversion, use \\
\midrule
\multirow{2}{*}{D5} & \multirow{2}{*}{Biochemical \& Food Processing} & D5m & \makecell{Industrial Biochemical Processes} & 4164 & production, fermentation, acid, lignin, hydrolysis, enzyme, activity, yield, process, concentration, ethanol, biomass, pretreatment, pulp, ph, laccase, medium, sugar, temperature, yeast \\
 & & D5n & \makecell{Plant Biochemistry \\ \& Food Science} & 11048 & activity, acid, oil, growth, production, content, extraction, food, protein, cell, plant, effect, biomass, gene, water, concentration, composition, fruit, quality, ltd \\
\midrule
\multirow{2}{*}{D6} & \multirow{2}{*}{Wastewater Treatment \& Soil Remediation} & D6o & \makecell{Wastewater \& \\ Waste Management} & 6316 & sludge, waste, production, treatment, soil, removal, wastewater, process, reactor, cod, digestion, biogas, energy, rate, nitrogen, methane, carbon, compost, biomass, concentration \\
 & & D6p & \makecell{Environmental Chemistry \\ \& Remediation} & 13595 & soil, removal, adsorption, water, concentration, treatment, ph, metal, process, cd, sludge, mg, cu, pb, wastewater, zn, extraction, degradation, cr, solution \\
\bottomrule
\end{tabular} 
}
\label{table2}
\end{table}

\subsubsection{Collaboration involving different research disciplines}
As shown thus far, the CE field is composed of a diverse array of research domains and researchers who have moved into the field from different backgrounds.
In this section, we examine whether interdisciplinary collaboration has emerged under the framework of Circular Economy by analyzing co-authorship as a proxy for collaboration.
We also evaluate the academic impact generated by such interdisciplinary collaborations.

Figure~\ref{fig5}(a) compares the number of co-authorships between disciplines during the decade before (2005–2014, left) and after (2015–2024, right) the EU’s first CE Action Plan.
Each color represents a researcher’s major discipline, and the width of the chords in the diagram corresponds to the volume of co-authorships.
Compared to the pre-2015 period, the post-2015 diagram shows a noticeable activation of collaborations across researchers from different original disciplines.
In particular, the chords connecting different fields are thicker after 2015, and new co-authorships have emerged in disciplinary areas that did not appear in the earlier period.
These observations suggest that interdisciplinary collaboration has been activated following the institutionalization of the CE field.

To compare areas where collaboration is occurring with those where it is not, we define two indicators:
(1) a normalized co-authorship intensity metric, which reflects the actual density of co-authorships observed between disciplines during the 2015–2024 period, and
(2) a semantic similarity metric, which represents the conceptual proximity originally shared between disciplines, based on the research focus prior to the institutionalization of the CE field.
For the normalized co-authorship intensity, we compute the Jaccard coefficient for each pair of researchers' original disciplines.
The Jaccard coefficient $J(D_i, D_j)$ is calculated using papers published between 2015 and 2024 that were co-authored by two or more CE active researchers. It is defined as:

\begin{equation}
J (D_i, D_j) = \frac{\text{(Number of co-author pairs between $D_i$ and $D_j$)}}{\text{(Number of co-author pairs including $D_i$ or $D_j$ in )}}
\end{equation}

This index normalizes co-authorship intensity between fields on a scale from 0 to 1.
For the semantic similarity metric, we first calculate the centroid vector of the semantic embeddings derived from the abstracts of papers published between 2005 and 2014 in each field.
Then, we compute the cosine similarity between these centroids to quantify the semantic similarity between research disciplines.

Figure~\ref{fig5}(c) plots all field pairs with their Jaccard coefficient on the y-axis and semantic similarity (cosine similarity) on the x-axis.
The color of each point indicates the median number of citations for papers resulting from those collaborations (red = highly cited, blue = less cited).

Semantic similarity was computed by calculating the centroid vectors of the abstract embeddings for each research discipline and measuring the cosine similarity between them.
The plot shows a positive correlation between semantic similarity and co-authorship intensity, especially among field pairs with cosine similarity above 0.8.

While this finding is intuitive, the data also reveal field pairs with similarly high semantic similarity but vastly different co-authorship levels.
For example, disciplines like (D1a: Business Management \& Innovation Studies and D1b: Systems \& Operations Management), and (D6o: Wastewater \& Waste Management and D6p: Environmental Chemistry \& Remediation) exhibit both high semantic similarity and strong co-authorship.

In contrast, discipline pairs such as (D4l: Renewable Energy \& Fuel Production and D6o: Wastewater \& Waste Management), (D2c: Environmental \& Agricultural Management and D2e: Sustainable Industrial Development), and (D5m: Industrial Biochemical Processes and D6o: Wastewater \& Waste Management) show high semantic similarity but low co-authorship.
These pairs may represent missed opportunities for interdisciplinary collaboration that could be fostered in future research initiatives.

Conversely, some discipline pairs show strong co-authorship despite relatively low semantic similarity.
For instance, (D3f: Material Science \& Nanotechnology and D4l: Renewable Energy \& Fuel Production) have a semantic similarity of approximately 0.7 but exhibit substantial co-authorship.
One of the most cited papers from this collaboration is ``A qualitative assessment of lithium ion battery recycling processes'' published in 2021 \citep{sommerville2021qualitative}.
Among its six authors, Dr. Mohammad Ali Rajaeifar was primarily active in biofuel research prior to 2014 (D4l: Renewable Energy \& Fuel Production) and began publishing on electric vehicles around 2019.
Meanwhile, Dr. Emma Kendrick (University of Birmingham) was originally active in materials chemistry and pigments research (D3f: Material Science \& Nanotechnology) in the mid-2000s and increasingly focused on lithium-ion batteries in the 2010s.
This paper exemplifies a successful collaboration between researchers from originally distinct disciplines.

Another example is the collaboration between fields (D1a: Business Management \& Innovation Studies and D2d: Sustainable Energy \& Building Systems).
Although their semantic similarity is also around 0.7, they exhibit high levels of co-authorship.
One of the most cited papers emerging from this collaboration is ``Sustainable supply chain management and the transition towards a circular economy: Evidence and some applications'' \citep{genoveseSustainableSupplyChain2017}, mentioned earlier.
Among its four authors, Dr. Adolf Acquaye (Khalifa University) and Dr. Lenny Koh (Sheffield University) originally belonged to D2d: Sustainable Energy \& Building Systems and D1a: Business Management \& Innovation Studies respectively, as discussed in the previous section.

So, is there a relationship between collaboration among researchers from different disciplines and the impact of their research?
In bibliometric studies, it has been reported that co-authorship between researchers with different disciplines tends to increase a paper's impact—as measured by citation counts \citep{wang_interdisciplinarity_2015}—and similar effects can be expected in the field of Circular Economy (CE).
To explore this, we analyzed whether citation counts differ between papers authored by \textit{intradisciplinary teams} and those authored by \textit{interdisciplinary teams}.
It is worth noting that many highly cited recent CE papers are review articles, a trend that has been critically highlighted \citep{dragomir_state_2024}.
Since researchers from different disciplines often collaborate to write review articles—and since these papers tend to receive disproportionately high citation counts—we excluded review articles (i.e., those with the word ``Review'' in the title or abstract) from our dataset to avoid citation inflation.
We further restricted our analysis to papers published in or before 2020, ensuring that each paper includes at least one active CE researcher among the authors. This allowed us to include only papers that had sufficient time to accumulate citations and enabled a more accurate evaluation of citation impact.

\begin{figure}[h]
\centering
\includegraphics[width=\linewidth]{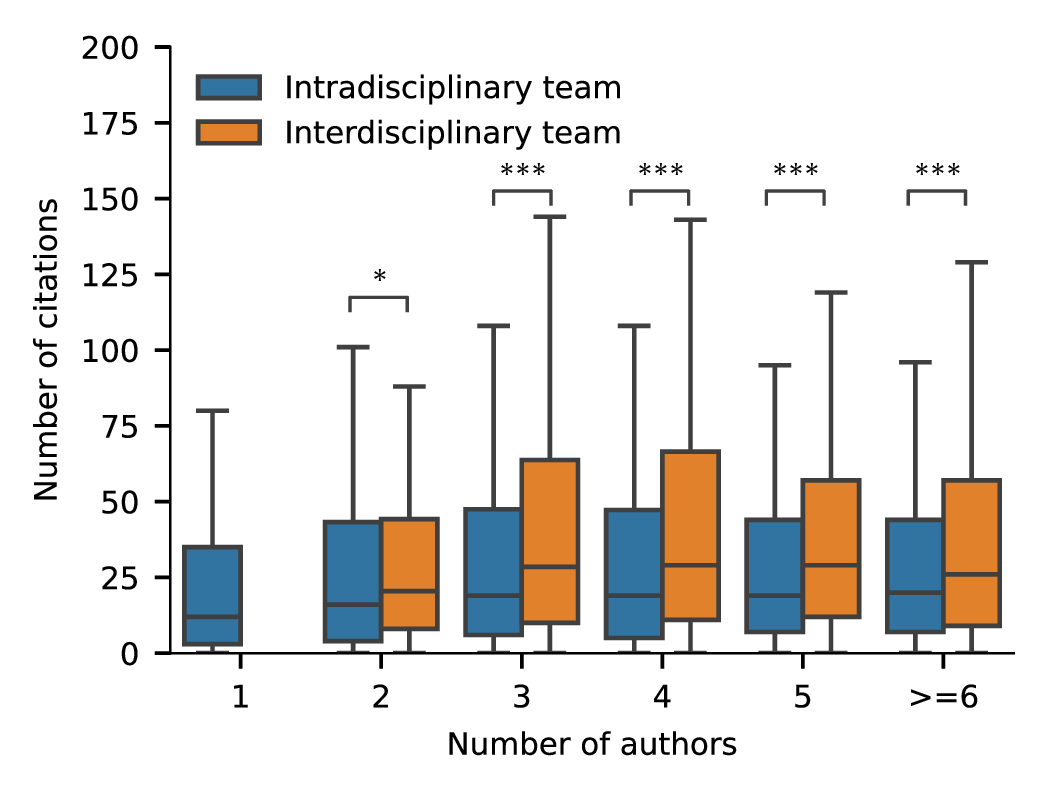}
\caption{Graph showing the relationship between citation count (y-axis) and number of authors (x-axis). Blue represents papers in which all active CE researchers among the authors share the same discipline, while orange represents papers in which the active CE researchers represent two or more distinct disciplines. A one-sided Mann--Whitney U test was conducted to compare the two groups. Asterisks indicate significance levels: *\,p\,$<$\,0.05, ***\,p\,$<$\,0.001.}
\label{fig6}
\end{figure}

Figure~\ref{fig6} displays citation counts (y-axis) plotted against the number of authors (x-axis), comparing papers authored by \textit{intradisciplinary teams} (blue) with those authored by \textit{interdisciplinary teams} (orange).
Here, an intradisciplinary team refers to a paper in which all active CE researchers among the authors share the same discipline, whereas an interdisciplinary team includes papers where the active CE researchers represent two or more distinct disciplines.
Note that for single-author papers, the author’s discipline is necessarily counted as one, so only the blue boxplot appears in that category.
The graph reveals that, regardless of the number of authors, papers written by interdisciplinary teams tend to receive higher citation counts.
This pattern remains consistent even when controlling for not only the number of authors, but also the number of active CE researchers included among them.
These results suggest that interdisciplinary collaboration among active CE researchers is positively associated with higher research impact within the Circular Economy domain.

\section{Conclusion \& Discussion}

Since the European Union released its Circular Economy (CE) Action Plan in 2015, the number of CE-related publications has increased rapidly.
The contribution of this paper lies in applying text embedding and clustering techniques to uncover the structure of this rapidly expanding interdisciplinary field, clarifying the attention it receives from academia and policy, and mapping collaboration networks among researchers.

Specifically, we addressed the following research questions: RQ1: Among the various research fields that constitute Circular Economy, which ones receive the most attention and citations from academic and policy perspectives? RQ2: Which types of interdisciplinary collaborations are emerging, and how are such collaborations related to research impact?

For RQ1, we classified the CE research landscape into six major and sixteen minor clusters, and analyzed the publication volume and citation impact from both academic and policy perspectives.
Our analysis revealed that business-related fields, particularly F1a: Business \& Supply Chain, attract considerable academic and policy attention.
The observed correlation between academic and policy citations is consistent with previous findings \citep{yinCoevolutionPolicyScience2021}.

Interestingly, a trade-off was observed between the magnitude of academic and policy attention and the amount of financial support across different research areas. 
While business-related fields attracted significant attention, they received relatively limited financial support (i.e., grants). 
Conversely, engineering-related fields received more financial support despite having fewer citations.
From a positive perspective, it is possible that research in the former areas draws attention to CE, while the latter areas attract financial support, creating a dynamic interaction between the two. 
In fact, since around 2016, a sharp increase in the amount of grant funding per paper has been observed in the latter fields. 
From a negative perspective, although the former fields are in the spotlight, the latter—where actual implementation of CE takes place—has not experienced the same growth in citations or publication volume. 
This may be due to a lower influx of researchers into the latter areas compared to the former. 
Such a situation raises concerns that a gap between scholarly/policy attention and real-world implementation may emerge in the future. 
For business-related fields, there is a need to propose more practical and applicable business models, while engineering-related fields must advance toward social implementation of technologies.

For RQ2, we showed that co-authorship between researchers from different disciplines has intensified since the announcement of the CE Action Plan in 2015.
Furthermore, papers co-authored by researchers from different disciplines tend to have higher impact, consistent with prior findings that diversity in author discipline is positively associated with long-term citations \citep{wang_interdisciplinarity_2015}.
The CE concept has provided a platform for interdisciplinary integration, which represents a significant contribution to academic discourse.

However, our analysis also revealed notable disparities in the extent of collaboration across research disciplines.
For instance, although D5m: Industrial Biochemical Processes and D6o: Wastewater \& Waste Management address closely related themes in wastewater treatment, their co-authorship rate is remarkably low.
Conversely, collaborations between researchers with business discipline (e.g., D1a: Business Management \& Innovation Studies) and those with engineering discipline (e.g., D2d: Sustainable Energy \& Building Systems) have yielded highly cited papers.
Such cases suggest that impactful, practice-oriented research may arise from stronger linkages between distinct disciplines, particularly between business and engineering domains.
Going forward, we anticipate that these collaborations will play an increasingly important role in bridging academic innovation with practical applications in industry.

\section*{Declaration of generative AI and AI-assisted technologies in the writing process}
During the preparation of this work the authors used ChatGPT in order to translate and check the grammar of our manuscript. After using this tool/service, the authors reviewed and edited the content as needed and take(s) full responsibility for the content of the published article.

\clearpage
\appendix

\section{Most cited paper in each research field.}

\begin{table}[H]
    \centering
    \resizebox{\textwidth}{!}{%
    \begin{tabular}{p{6cm}p{8cm}p{8cm}p{4cm}cc}
\toprule
\textbf{Field Name} & \textbf{Title} & \textbf{Authors} & \textbf{Journal} & \textbf{Year} & \textbf{Cited Number} \\
\midrule
F1a: Business \& Supply Chain & The Circular Economy - A new sustainability paradigm? & M. Geissdoerfer, P. Savaget, N. M. P. Bocken, E. J. Hultink & Journal of Cleaner Production & 2017 & 3984 \\
F1b: Fashion \& Textile & Environmental impact of textile reuse and recycling - A review & G. Sandin, G. M. Peters & Journal of Cleaner Production & 2018 & 535 \\
F1c: Manufacturing \& Product Design & Consumer product knowledge and intention to purchase remanufactured products & Y. Wang, B. T. Hazen & International Journal of Production Economics & 2016 & 298 \\
F2d: Waste Management \& Recycling & Construction and demolition waste management in China through the 3R principle & B. Huang, X. Wang, H. Kua, Y. Geng, R. Bleischwitz, J. Ren & Resources, Conservation and Recycling & 2018 & 623 \\
F2e: Construction \& Urban Development & Circular economy for the built environment: A research framework & F. Pomponi, A. Moncaster & Journal of Cleaner Production & 2017 & 567 \\
F2f: Resource Management & The: E factor 25 years on: The rise of green chemistry and sustainability & R. A. Sheldon & Green Chemistry & 2017 & 944 \\
F3g: Biomass Energy \& Waste Management & Green, circular, bio economy: A comparative analysis of sustainability avenues & D. D'Amato, N. Droste, B. Allen, M. Kettunen, K. Lähtinen, J. Korhonen, P. Leskinen, B.D. Matthies, A. Toppinen & Journal of Cleaner Production & 2017 & 644 \\
F3h: Food Processing \& Valorization & Health effects of phenolic compounds found in extra-virgin olive oil, by-products, and leaf of olea europaea L. & A. Romani, F. Ieri, S. Urciuoli, A. Noce, G. Marrone, C. Nediani, R. Bernini & Nutrients & 2019 & 250 \\
F3i: Agriculture \& Waste Reuse & The potential implications of reclaimed wastewater reuse for irrigation on the agricultural environment: The knowns and unknowns of the fate of antibiotics and antibiotic resistant bacteria and resistance genes - A review & A. Christou, A. Agüera, J. M. Bayona, E. Cytryn, V. Fotopoulos, D. Lambropoulou, C. M. Manaia, C. Michael, M. Revitt, P. Schröder, D. Fatta-Kassinos & Water Research & 2017 & 419 \\
F4j: Industrial Manufacturing \& Recycling & Exploring Industry 4.0 technologies to enable circular economy practices in a manufacturing context: A business model proposal & D. L. M. Nascimento, V. Alencastro, O. L. G. Quelhas, R. G. G. Caiado, J. A. Garza-Reyes, L. Rocha-Lona, G. Tortorella & Journal of Manufacturing Technology Management & 2019 & 560 \\
F4k: Plastic Production \& Recycling Management & Beyond Mechanical Recycling: Giving New Life to Plastic Waste & I. Vollmer, M. J. F. Jenks, M. C. P. Roelands, R. J. White, T. Harmelen, P. Wild, G. P. Laan, F. Meirer, J. T. F. Keurentjes, B. M. Weckhuysen & Angewandte Chemie - International Edition & 2020 & 862 \\
F5l: Civil Engineering Materials & Use of recycled aggregates arising from construction and demolition waste in new construction applications & R.V. Silva, J. de Brito, R.K. Dhir & Journal of Cleaner Production & 2019 & 209 \\
F6m: Wastewater Treatment & The Global Food-Energy-Water Nexus & P. D'Odorico, K. F. Davis, L. Rosa, J. A. Carr, D. Chiarelli, J. Dell’Angelo, J. Gephart, G. K. MacDonald, D. A. Seekell, S. Suweis, M. C. Rulli & Reviews of Geophysics & 2018 & 431 \\
F6n: Metal Adsorption \& Recovery & Removal of chromium from wastewater by membrane filtration, chemical precipitation, ion exchange, adsorption electrocoagulation, electrochemical reduction, electrodialysis, electrodeionization, photocatalysis and nanotechnology: a review & H. Peng, J. Guo & Environmental Chemistry Letters & 2020 & 346 \\
F6o: Metals \& Mining Industry & Recovery and recycling of lithium: A review & B. Swain & Separation and Purification Technology & 2017 & 1103 \\
F6p: Lithium-Ion Battery Technology & Sustainability and in situ monitoring in battery development & C. P. Grey, J. M. Tarascon & Nature Materials & 2016 & 947 \\
\bottomrule
\end{tabular} 
    \label{tab:appendix_a}
    }
\end{table}

\bibliographystyle{elsarticle-harv} 
\bibliography{references}



\end{document}